\documentclass[prd,aps,showpacs,showkeys,twocolumn,10pt]{revtex4-1}
\usepackage{amssymb,amsmath,amsthm}
\usepackage{mathrsfs}
\usepackage[applemac]{inputenc}

\def\widebar{\overline}

\begin{document}

\title{Horndeski theories self-tuning to a de Sitter vacuum.}

\author{Prado Mart\'in-Moruno}
\email{pmmoruno@fc.ul.pt}
\author{Nelson J.~Nunes}
\email{njnunes@fc.ul.pt}
\author{Francisco S.~N.~Lobo}
\email{fslobo@fc.ul.pt}
\affiliation{Instituto de Astrof\'isica e Ci\^encias do Espa\c{c}o, Universidade
 de Lisboa, Faculdade de Ci\^encias, Campo Grande, PT1749-016 Lisboa, Portugal}

\date{\today}

\begin{abstract}
We consider Horndeski cosmological models able to screen the vacuum energy coming from any field theory assuming that after this screening the space should be in a de Sitter vacuum with a particular value 
of the cosmological constant specified by the theory of gravity itself. The most general scalar-tensor cosmological models without higher than second order derivatives in the field equations that have a spatially flat de Sitter critical point for \emph{any} kind of material content or vacuum energy are, therefore, presented. These models could allow us to understand the current accelerated expansion of  the universe as the result of  a dynamical evolution towards a de Sitter attractor.
\end{abstract}


\maketitle

\section{Introduction}
The possibility of understanding the late-time accelerated expansion of the Universe without invoking the cosmological constant problem has revived the interest on alternative theories of gravity. 
Although most modified theories of gravity are still affected by this problem if the vacuum energy gravitates \cite{Barcelo:2014qva}, their rich structure could allow us to describe the cosmic speed-up 
by assuming that the vacuum energy must be fixed to zero. Therefore, more than a decade after the discovery of the late-time cosmic acceleration, we are essentially facing the same questions that 
were considered after the formulation of general relativity, regarding its uniqueness and the consistency of its rival theories.
Some progress, however, has been achieved by the Horndeski models. 
The Lagrangian for these models was first written down by Horndeski in 1974 \cite{Horndeski:1974wa} and recently rediscovered by Deffayet {\it et al.} \cite{Deffayet:2011gz}, when generalizing the covariantization 
for curved spacetimes \cite{Deffayet:2009wt} of the galileon models \cite{Nicolis:2008in}. 
In the galileon framework one allows general nonlinear derivative interactions for the Lagrangian of the scalar field which lead to second order field equations  \cite{deRham:2014zqa}.  
Horndeski theories are a generalization of the latter and are, therefore, free of the Ostrogradski instability. 
Note that the Horndeski Lagrangian contains four arbitrary functions of the scalar field and its kinetic energy, thus a complete general study of its cosmological implications is expected to be a difficult undertaking. 

The form of the Horndeski Lagrangian can be significantly simplified by requiring self-tuning properties, as has been done by
Charmousis {\it et al.}~\cite{Charmousis:2011bf,Charmousis:2011ea}.
This self-tuning consists on the screening of any cosmological constant by the scalar field leading to a Minkowski vacuum space, 
which avoids potential issues with field theory and is compatible with unitarity of the S matrix.
Thus, these models could alleviate the cosmological constant problem (at least from a classical point of view), as the field is not only able to overcome a high energy cosmological constant
but it also cancels its effect independently of its value or any variations on it, i.e., independently of the occurrence of a phase transition that could change the value
of the vacuum energy by a finite amount \cite{Charmousis:2011bf,Charmousis:2011ea}.
The resulting models, which have a Minkowski critical point for any material content, are characterized by four different potentials that were denoted the {\it fab four} \cite{Charmousis:2011ea}. 
The cosmology of these models was studied in Ref.~\cite{Copeland:2012qf}
considering no material content, where it was shown that the fab four potentials produce an effect similar to considering stiff matter, radiation or curvature in cosmological scenarios.

More recently, a nonlinear generalization of some fab four terms resulted in the so-called {\it fab five} 
models  \cite{Appleby:2012rx}. 
This nonlinear combination of purely kinetic gravity terms can give rise to an accelerating universe without the addition of extra propagating degrees of freedom on cosmological 
backgrounds when matter and radiation are present \cite{Appleby:2012rx}. 
The concept of self-tuning was extended to models which evolve toward a de Sitter vacuum instead of Minkowski, which also bring a large cosmological constant under control, though only for a particular material 
content for the universe. 
Nevertheless, in the fab five models the screening may be too effective to ensure stable cosmologies \cite{Linder:2013zoa}.

At a first glance one could think that the fast screening of the fab five \cite{Linder:2013zoa} together with 
the apparent difficulty of obtaining late time accelerating cosmological solutions 
of the fab four potentials \cite{Copeland:2012qf} 
may lead to the impossibility of describing a consistent cosmology alleviating the cosmological constant problem through a self-tuning 
mechanism. However, if the Horndeski cosmological models could deliver a de Sitter attractor instead of a Minkowski one (as in the fab five case) for any material content (as in the fab four models), the universe
would naturally approach a late-time phase of accelerated expansion which would not be driven by the vacuum energy of the underlying particle theory.

In the present article we investigate the Horndeski models that are able to screen the vacuum energy coming from any field theory to a de Sitter vacuum. 
Although one loses the interesting aspects of the Minkowski space present in the fab four, here we are more interested in constructing a viable cosmology compatible with current observations and independent of the vacuum energy.

The paper is outlined as follows: In Sec.~\ref{generalF}, we extend the concept of self-tuning to the case in which one is left with a non-Minkowski geometry after the screening takes place;
therefore, we summarize some results presented in 
Ref.~\cite{Charmousis:2011bf,Charmousis:2011ea} generalizing them for any spatially flat critical point for cosmological scenarios.
In Sec.~\ref{Requiring}, we focus our attention to a spatially flat de Sitter critical point and obtain two families of solutions able to self-tune to this vacuum. 
In Sec.~\ref{discussion}, we discuss the results of this article.
Finally, we relegate to App.~\ref{diccionario} the definition of the Horndeski Lagrangian and its relation with the form introduced by Deffayet {\it et al.}~\cite{Deffayet:2011gz},
and to App.~\ref{funcionesD} the functions corresponding to the linear models in the notation of Ref.~\cite{Deffayet:2011gz}.

\section{General Framework}\label{generalF}
The Horndeski point-like Lagrangian with a scalar field $\phi$ can be written as \cite{Charmousis:2011ea}
\begin{equation}\label{Lsimple}
L_{\rm H}=a^3\sum_{i=0}^3\left[X_i\left(\phi,\,\dot\phi\right)-\frac{k}{a^2} Y_i\left(\phi,\,\dot\phi\right)\right]\,H^i,
\end{equation}
where $a$ is the scale factor, $H=\dot a/a$ is the Hubble parameter, a dot represents a derivative with respect the cosmic time $t$, $k$ is the spatial curvature, 
and the functions $X_i$ and $Y_i$ can be written in terms of the usual Horndeski functions 
as shown in Ref.~\cite{Charmousis:2011ea} (see App.~\ref{diccionario}).
We assume that the material content is minimally coupled to gravity and uncoupled to the field $\phi$. Thus, the matter minisuperspace Lagrangian is
\begin{equation}\label{Lm}
L_{\rm m}=-a^3\rho_{\rm m}(a),
\end{equation}
where the material satisfies the usual conservation equation $\dot\rho_{\rm m}+3H(1+w_{\rm m})\rho_{\rm m}=0$, and $w_{\rm m}=p_{\rm m}/\rho_{\rm m}$ is the equation of state parameter.

In order to obtain the most general model capable of screening any vacuum energy into a Minkowski vacuum, Charmousis {\it et al.} showed that one needs to impose the following three conditions:
$(i)$ the on-shell minisuperspace Lagrangian must be independent of the field and its derivatives (up to total derivatives);
$(ii)$ the on-shell Hamiltonian density must depend on $\dot\phi$ and; 
$(iii)$ the full scalar equation of motion must be dependent on $\ddot a$. 
The term on shell simply refers to evaluation at the cosmological solution which arises after complete screening takes place.
The two first conditions imply that the field equation is trivially satisfied once screening has occurred, as it is needed to have a screening mechanism effective for any value of the cosmological constant.
Since we are assuming that matter is not interacting with the field $\phi$, the field equation is not only independent of the vacuum energy but also of the material content.
This implies that complete screening is reached independently of the material content given by Lagrangian (\ref{Lm}), as pointed out in Ref.~\cite{Charmousis:2011ea}.
The way in which the field screens the vacuum energy and this matter content, however, does depend on the particular material present in the theory through the modified Friedmann equation. 
The existence of non-trivial cosmological solutions which depend on the matter content is allowed by condition $(iii)$, which is the requirement for the self-tuning solution to be approached dynamically. More specifically, the on shell solution is a critical point of the evolution.

A similar procedure as that presented in Ref.~\cite{Charmousis:2011ea} can be followed considering now that after the screening effect the universe has an on shell solution in the scale factor, 
$a_{\rm os}$, characterizing the vacuum. 
Thus,  the model can have a critical point at $H_{\rm os}$ where the subscript ``os'' denotes evaluation on shell in $a$ (which is Minkowski in \cite{Charmousis:2011ea}).
The first condition implies that the Lagrangian density evaluated on shell, $\mathcal{L}_{\rm os}$, can be written as
\begin{equation}\label{c1}
\mathcal{L}_{\rm os}=c(a_{\rm os})+\frac{1}{a_{\rm os}^3}\dot\zeta\left(a_{\rm os},\,\phi\right).
\end{equation}
where $c$ and $\zeta$ are arbitrary functions.
Conditions $(ii)$ and $(iii)$ are equivalent if $H_{\rm os}\neq0$ and imply that at least for one $i\neq0$, one has
 \begin{equation}\label{c2}
  X_{i,\dot\phi}\left(\phi,\,\dot\phi\right)-\frac{k}{a_{\rm os}^2} Y_{i,\dot\phi}\left(\phi,\,\dot\phi\right)\neq0.
 \end{equation}

Let us now consider three aspects with the aim of finding the general form of the functions $X_i$ which lead to self-tuning.
First, one must realise that a particular Lagrangian which satisfies the conditions given by Eqs. (\ref{c1}) and (\ref{c2}) for $k=0$ (which implies removing the dependence on the $Y_i$ functions) is given by
 \begin{equation}\label{Lc}
 \widebar L=a^3\left\{c(a)+\sum_{i=0}^3\widebar X_{i}\left(\phi,\,\dot\phi\right)\left[H^i-H_{\rm os}^i\right]\right\},
\end{equation}
which is a trivial generalization of that presented in Ref.~\cite{Charmousis:2011ea} for a (at the moment unspecified) critical point at $H_{\rm os}$. At the same time (\ref{Lc}) is more restricted because
it is valid only for $k=0$. Secondly, it should be noted that, 
in principle this is not the only Lagrangian with such a behaviour. 
As it has been explicitly proven in~\cite{Charmousis:2011ea} (with a proof valid for any $H_{\rm os}(a)$, i.e., only dependent of the scale factor), 
two Horndeski theories which self-tune to $H_{\rm os}$ are related through a total derivative of a function $\mu\left(a,\,\phi\right)$. 
Therefore, the general form of a Lagrangian able to self-tune is given by 
\begin{equation}\label{Ls}
 L=\widebar L+\dot\mu\left(a,\,\phi\right).
\end{equation}
A total derivative does not affect, of course, the cosmological evolution. 
The consideration of Eq.~(\ref{Ls}), however, allows us to restrict the general form of the $X_i$'s appearing in Lagrangian (\ref{Lsimple}) able to self-tune, by considering
their relation with $\widebar X_i$. 
Thirdly, given that $L$ and $\widebar L$ in this relation are given by Eqs.~(\ref{Lsimple}) and (\ref{Lc}), respectively, and that 
this relation must be satisfied during the whole evolution (that is, for any $H$), the particular form of the functions $X_i$ in terms of $\widebar X_i$ can be obtained. 
They are
\begin{equation}\label{Z0}
 X_0=c(a)-\sum_{i=1..3}\widebar X_{i}H_{\rm on}^i+\frac{\dot\phi}{a^3}\mu_{,\phi},
\end{equation}
\begin{equation}\label{Zi}
 X_1=\widebar X_1+\frac{1}{a^2}\mu_{,a},\qquad X_2=\widebar X_2,\qquad X_3=\widebar X_3.
\end{equation}
Combining these expressions to eliminate the dependence in $\widebar X_i$, it results that 
\begin{equation}\label{Lon}
 \sum_{i=0}^3 X_i(\phi,\dot\phi) H_{\rm os}^i=c(a)+\frac{H_{\rm os}}{a^2}\mu_{,a}+\frac{\dot\phi}{a^3}\mu_{,\phi},
\end{equation}
where the LHS is precisely the on-shell Lagrangian density $\cal{L}_{\rm os}$.

\section{Requiring a spatially flat de Sitter critical point}\label{Requiring}
Let us now focus our attention on Horndeski cosmologies with a spatially flat de Sitter critical point, which means that we demand $H_{\rm os}=\sqrt{\Lambda}$. As the LHS of  Eq.~(\ref{Lon}) is independent of $a$, the RHS must also be independent of $a$ for any value of $\dot\phi$. Keeping this in mind, we can constrain the general form of the function $\mu$ and, consequently, 
the form of the Lagrangian density evaluated on shell in $a$. This is provided by
\begin{equation}\label{Lonshell}
 \mathcal{L}_{\rm os}=3\sqrt{\Lambda}\,h(\phi)+\dot\phi\, h_{,\phi}(\phi),
\end{equation}
where $h(\phi)$ is a general function of $\phi$. Therefore, there are two different kinds of terms which can appear in the Lagrangian. 
These are: $(i)$ $X_i$--terms linear in $\dot\phi$, and 
$(ii)$ $X_i$--terms with a non-linear dependence on $\dot\phi$ in the Lagrangian (\ref{Lsimple}), but
which contribution vanishes in the on-shell point-like Lagrangian (\ref{Lonshell}).\\

\subsection{Terms linear in $\dot\phi$}
 
In order to satisfy Eq.~(\ref{Lonshell}) considering
only terms linear in $\dot\phi$, it is sufficient to consider
\begin{equation}\label{lineargral}
 X_i^{\rm linear}\left(\phi,\,\dot\phi\right)=3\sqrt{\Lambda} \,U_i(\phi)+\dot\phi\,W_i(\phi).
\end{equation}
As from Eq.~(\ref{Lonshell}) one has $\sum_{i=0}^3U_i(\phi)\Lambda^{i/2}=h(\phi)$ and $\sum_{i=0}^3W_i(\phi)\Lambda^{i/2}=h_{,\phi}(\phi)$,
the potentials $U_i$ and $W_i$ must satisfy the constraint
\begin{equation}\label{condnl}
  \sum_{i=0}^3W_i(\phi)\Lambda^{i/2}=\sum_{i=0}^3U_{i,\phi}(\phi)\Lambda^{i/2}.
\end{equation}
Thus, the Lagrangian of these models can be written as
\begin{equation}\label{Lglinear}
 L_{\rm linear}=a^3 \sum_{i=0}^3\left[3\sqrt{\Lambda} \,U_i(\phi)+\dot\phi\,W_i(\phi)\right]H^i.
\end{equation}
Taking into account this Lagrangian, the field equation can now be obtained. This is given by 
$a^3 \varepsilon_{\rm linear} \equiv  \partial{L}/\partial \phi - \partial_t (\partial L/\partial \dot{\phi})  =0$, which yields
\begin{eqnarray}\label{fieldg}
   \varepsilon_{\rm linear}&=&\sum_{i=0}^3 \left\{ 3\sqrt{\Lambda} \left[U_{i,\phi}(\phi)-\frac{H} {\sqrt{\Lambda}}W_i(\phi)\right]     \right. \nonumber\\
    &&   \left.  -i  \frac{\dot H}{H} W_i(\phi)  \right\}  H^i  \,.
\end{eqnarray}
As one expects, this field equation is independent of $\dot\phi$ and $\ddot\phi$, given that we have a Lagrangian linear on $\dot\phi$.
Taking into account condition (\ref{condnl}), it can be noted that Eq.~(\ref{fieldg}) is trivially satisfied for $\dot H=0$ and $H=\sqrt{\Lambda}$, 
independently of value of the field, any potential vacuum energy and material content, as we intended to.
This ensures, at the end of the day, the fulfilment of condition $(i)$.
We further require that the first term of Eq.~(\ref{fieldg}) cannot vanish for any $H$,
otherwise $\dot H$ would have to vanish during the entire evolution, which means that the screening would not be dynamical.
Furthermore, condition $(iii)$ is satisfied for $\sum_{i=0}^3iW_i\neq0$. 

The Hamiltonian density is given by
\begin{equation}\label{Hglinear}
 \mathcal{H}_{\rm linear}=\sum_{i=0}^3\left[3(i-1)\sqrt{\Lambda}\,U_i(\phi)+i\,\dot\phi\,W_i(\phi)\right]H^i \,,
\end{equation}
and the modified Friedmann equation is given by $\mathcal{H}_{\rm linear}=-\rho_m(a)$.
Here the material content first appears, affecting the cosmological evolution only until the field completely screens its contribution.
Condition $(ii)$ requires that this Hamiltonian density evaluated on shell depends on $\dot\phi$ to absorb any changes on the value of the vacuum energy and the matter in the field, 
which implies $\sum_{i=0}^3iW_i\neq0$ as condition $(iii)$.
Therefore, a necessary and sufficient condition for the linear models to self-tune is that at least one $W_i\neq0$, with $i\neq0$ is present. 
The models containing only $W_0$ and $U_i$ potentials and those with only $U_i$'s do not spoil the screening of other potentials when combined with them, but they are not able to self-tune by themselves (namely, to approach a de Sitter solution dynamically). 
It must be noted that an Einstein--Hilbert term is contained in the Lagrangian of Eq.~(\ref{Lglinear}), although this term is not  able to self-tune by itself. It can be explicitly written by redefining the potential
$U_2$ as $U_2-1/(8\pi G\sqrt{\Lambda})$.

Considering the relation of the functions $X_i$ and $Y_i$ with the Horndeski functions given in Ref.~\cite{Charmousis:2011ea}  (and summarized in App.~\ref{diccionario}), 
one can obtain those functions which will give the Horndeski Lagrangian without restriction to the minisuperspace. These are given by
\begin{eqnarray}
 \kappa_1(\phi,\,\dot\phi) &=&\frac{3\sqrt{\Lambda}\, U_3}{4\sqrt{-X}}
 -\frac{W_{3}}{8}\,{\rm ln}(-X),\label{kappa1} \\
 \kappa_3(\phi,\,\dot\phi)&=&\frac{U_4}{\sqrt{-X}}+U_5 +\frac{W_{3,\phi}}{8}\left[2-{\rm ln}(-X)\right]\nonumber\\
 &&+ \frac{(9\sqrt{\Lambda}U_{3,\phi}-W_{2})\,{\rm ln}(-X)}{24\sqrt{-X}},\label{kappa3}\\
  \kappa_8(\phi,\,\dot\phi)&=&\frac{3\sqrt{\Lambda}\,U_{2,\phi}-W_{1}}{6 X}
  -\frac{W_{2,\phi}}{3\sqrt{-X}},\label{kappa8} \\
 \kappa_9(\phi,\,\dot\phi)&=&3\sqrt{\Lambda}U_0+\left[W_{0}-\sqrt{\Lambda}U_{1,\phi}\right]\sqrt{-X}\nonumber\\
 &&+\frac{W_{1,\phi}X}{2}+2U_{5,\phi\phi}X^2,\label{kappa9} \\
 F(\phi,\,\dot\phi)&=&-\frac{\sqrt{\Lambda}\,U_2}{4}-\frac{W_{2}\sqrt{-X}}{6}-U_5X, \label{Ffinal}
\end{eqnarray}
where $X=\nabla_\mu\phi\nabla^\mu\phi$, $U_4(\phi)$ and $U_5(\phi)$ are integration constants with respect to $X$, and the dependence of the potentials on the field is implicitly assumed in order to simplify notation.
Two additional functions which appeared as integration constants with respect to $X$ have been removed since their contributions to the different functions lead to a total derivative for the general Horndeski 
Lagrangian. As the potentials $U_4(\phi)$ and $U_5(\phi)$ do not appear in the minisuperspace Lagrangian for spatially flat models, these terms are not only  unable of screening, but they also do not affect 
the cosmological dynamics for cosmologies with $k=0$. Nevertheless, one should take into account their possible existence when considering solutions of these models with a different symmetry. 
On the other hand, in App.~\ref{funcionesD} we include the functions that appear in the Lagrangian when written as in Ref.~\cite{Deffayet:2011gz}.

In summary, for this class of solutions, we have ten functions of $\phi$ appearing in the Horndeski Lagrangian, $U_i(\phi)$ with $i=0,...,5$ and $W_j(\phi)$ with $j=0,...,3$, and one constraint, given by Eq.~(\ref{condnl}).
Therefore, we are left with nine arbitrary potentials, which reduces to seven arbitrary 
potentials when we restrict ourselves to the study of Friedmann-Robertson-Walker models, where $U_4(\phi)$ and $U_5(\phi)$ are not present.

\subsection{Nonlinear terms}
 
Let us now consider terms with an arbitrary dependence on $\phi$ and $\dot\phi$, leading to a Lagrangian given by
\begin{equation}\label{Lnl}
 L_{\rm nl}=a^3 \sum_{i=0}^3X^{\rm nl}_i\left(\phi,\,\dot\phi\right)\,H^i.
\end{equation}
Taking into account Eq.~(\ref{Lonshell}), we know that the contribution to the point-like Lagrangian of these terms vanishes 
on shell in $a$. As each $X_i$ is multiplied by a different power of $H$ in the Lagrangian (\ref{Lsimple}), in order to have an arbitrary function of $\phi$ and $\dot\phi$
appearing in one particular $X_m$, that function must cancel its contribution only on shell in Eq. (\ref{Lonshell}) with the other functions $X_i$ with $i\neq m$. Thus, the second family of solutions contain terms 
$X^{\rm nl}_i$ in the Lagrangian that can be nonlinear in $\dot\phi$ and satisfy the constraint
\begin{equation}\label{condnnl}
 \sum_{i=0}^{3} X^{\rm nl}_i\left(\phi,\,\dot\phi\right)\Lambda^{i/2}=0.
\end{equation}
These models have a Hamiltonian density 
\begin{equation}\label{Hnlg}
 \mathcal{H}_{\rm nl}=\sum_{i=0}^{3}\left[(i-1)X^{\rm nl}_i\left(\phi,\,\dot\phi\right)+\dot\phi \,X^{\rm nl}_{i,\dot\phi}\left(\phi,\,\dot\phi\right)\right]H^i,
\end{equation}
which does not vanish on shell in general, since we have
\begin{equation}
 \mathcal{H}_{\rm nl,\,os}=\sum_{i=0}^{3}i\, X^{\rm nl}_i\left(\phi,\,\dot\phi\right)\Lambda^{i/2},
\end{equation}
where we used Eq.~(\ref{Hnlg}) on shell and constraint (\ref{condnnl}) together with its first derivative.
As the Hamiltonian density (\ref{Hnlg}) depends on $\dot\phi$ in general, one can screen different vacuum energies to a de Sitter solution fixed by the theory even if the on-shell minisuperspace Lagrangian vanishes, 
that is, even if $h(\phi)=0$ in Eq. (\ref{Lonshell}). 
It should be noted that, if only these nonlinear terms are present, the modified Friedmann equation is $\mathcal{H}_{\rm nl}=-\rho_m(a)$, but if both kind of terms appear in the Lagrangian one has
$\mathcal{H}_{\rm linear}+\mathcal{H}_{\rm nl}=-\rho_m(a)$.
On the other hand, the field equation can be written as 
\begin{eqnarray}\label{fieldnong}
 \varepsilon_{\rm nl}&=&\sum_{i=0}^{3}   \left[X^{\rm nl}_{i,\phi}-3X^{\rm nl}_{i,\dot\phi}H- iX^{\rm nl}_{i,\dot\phi}\frac{\dot H}{H}
 \right.\nonumber\\
 &&- \left.
 X^{\rm nl}_{i,\dot\phi \phi}\dot\phi-X^{\rm nl}_{i,\dot\phi\dot\phi}\ddot\phi\right] H^i.
\end{eqnarray}
From this equation one can explicitly verify that it vanishes identically for $\{H=\sqrt{\Lambda},\,\dot H=0\}$, independently of the value of $\phi$ or $\rho_{\rm m}$,
due to condition (\ref{condnnl}) and other conditions which can be obtained deriving condition (\ref{condnnl}) with respect to $\phi$ or $\dot\phi$.
Again cases with Eq.~(\ref{fieldnong}) reducing to an algebraic equation in $H$ have to be dismissed.

From the nonlinear family of models, the ones with only a pair of functions $X^{\rm nl}_{i}$ must be emphasized as they have a particular simple expression.
Since these terms have to appear in pairs, they can be written as
\begin{equation}\label{Xp}
  X^{\rm p}_i\left(\phi,\,\dot\phi\right)=\Lambda^{-i/2}\sum_{j=0}^3{\rm sgn}(j-i)\,W_{ij}\left(\phi,\,\dot\phi\right), 
\end{equation}
where $W_{ij}=W_{ji}$, $W_{ij,\dot\phi\dot\phi}\neq 0$, and ${\rm sgn}(j-i)$ 
is the sign function.
Thus, there are six arbitrary functions of $\phi$ and $\dot\phi$ appearing in the Lagrangian. This non-linear Lagrangian is 
\begin{equation}\label{Lnonlinear}
 L_{\rm p}=a^3\sum_{i,j=0}^{3}{\rm sgn}(j-i)W_{ij}\left(\phi,\,\dot\phi\right)\left(\frac{H}{\sqrt{\Lambda}}\right)^i,
\end{equation}
which, of course, vanishes on-shell. The Hamiltonian density and field equation for these models are given by Eqs.~(\ref{Hnlg}) and (\ref{fieldnong}) substituting $X^{\rm nl}_{i}$ by
$X^{\rm p}_i$ defined in Eq.~(\ref{Xp}).

The nonlinear family of models is too general to write the Horndeski functions without restricting attention to a particular group of models. 
In order to compare these models to other existing in the literature the simplest approach is to proceed at the level of the minisuperspace Lagrangian.

\subsection{Most general Lagrangian}
There is still a group of terms that are not yet included in our study. These terms contribute to the point-like Lagrangian with $k=0$ via a total derivative, 
which do not necessarily imply that these additional terms are equivalent to a total derivative in the general Horndeski Lagrangian (before restricting to the minisuperspace); 
they are not even a total derivative for the minisuperspace Lagrangian for $k\neq0$ (for which no de Sitter critical point is in principle present). 
Therefore, the most general point-like Lagrangian with a spatially flat de Sitter critical point for any material content and vacuum energy, and which equations of motion do not contain 
higher than second derivative is given by
\begin{equation}\label{Ltodo}
 L=L_{\rm linear}+L_{\rm nl}+\frac{{\rm d}}{{\rm d}t}F(a,\,\phi)+k\,G(a,\,\dot a,\,\phi,\,\dot\phi),
\end{equation}
with $L_{\rm linear}$ and $L_{\rm nl}$ given by Eqs.~(\ref{Lglinear}) and (\ref{Lnl}), respectively.
These additional terms are, of course, unable to screen by themselves, as the cosmological dynamics of spatially flat models is independent of them.

A particular example of a model described by a Lagrangian of the form (\ref{Ltodo}) has been studied in Ref.~\cite{Gubitosi:2011sg}. Their ``purely kinetic coupled gravity'' corresponds to a model
with
\begin{equation}
 U_2=-\frac{M_{\rm Pl}^2}{\sqrt{\Lambda}},\,\,\,\, X_0^{\rm nl}=\frac{\dot\phi^2}{2},\,\,\,\,X_2^{\rm nl}=-\frac{\dot\phi^2}{2\Lambda}
\end{equation}
and the remaining potentials vanish ($c_3=-M_{\rm Pl}^2/(6\Lambda)$ in their notation). This model has, therefore, a de Sitter critical point for any value of the vacuum energy and kind of material content.

\section{Summary and further comments}\label{discussion}
In this article we have extended the concept of self-tuning presented in Refs.~\cite{Charmousis:2011bf,Charmousis:2011ea} to a de Sitter vacuum solution. 
This self-tuning is based on a field equation with an explicit dependence on the derivative of the Hubble parameter and that is trivially satisfied at the critical point. 
This feature allows the field to dynamically screen any vacuum energy through the modified Friedmann equation, but also any
other material which does not interact with the field.
Thus, requiring that the Horndeski minisuperspace Lagrangian has to satisfy this self-tuning to de Sitter, we have obtained two families of cosmological scenarios with a spatially flat de Sitter critical point.
The first family has a minisuperspace Lagrangian consisting of  linear terms in $\dot\phi$, and the second family has a vanishing on-shell minisuperspace Lagrangian.

We have generalized  a powerful framework that has helped to select the models which one should take into account to alleviate (at least classically) the cosmological constant problem.
These models are phenomenological promising because in the case when the critical point is indeed an attractor, these cosmologies allow a de Sitter late-time accelerated phase independent of the particular value of the vacuum energy.
Thus, these models allow us to understand the late-time cosmic speed-up as the result of 
a dynamical approach to a critical fixed point in field space.

The stability analysis of the critical point, however, has to be undertaken given a particular model as in the fab four case \cite{Copeland:2012qf}.
It is also now important to study the evolution of the Universe under these models during radiation and matter phases in the search for viable cosmological models leading to a late-time accelerated expansion of the 
Universe compatible with current observations. 
In Ref.~\cite{Martin-Moruno:2015lha} we have already concluded that the critical point is indeed an attractor for relevant models of the linear family, 
showing that those models have a promising behaviour at the cosmological background level.
Work along these lines regarding the nonlinear family is currently in progress.

It must be emphasized that the models presented in this article are generically different from other models obtained in the literature \cite{Heisenberg:2014kea, DeFelice:2010pv}.
Those models have a de Sitter attractor for particular kinds of Horndeski theories when a given material content for the universe is assumed \cite{Heisenberg:2014kea, DeFelice:2010pv}. 
The advantage of our model is that it may alleviate the cosmological constant problem because it has a de Sitter critical point for {\it any} value of the vacuum energy and type of material 
content.  This implies that the cosmological constant of the critical point is not determined by the energy of the vacuum of the matter field but is of purely gravitational origin.

Finally, it must be noted that we are only considering Horndeski models. There could be in principle other stable theories with a self-tuning mechanism which do not have second order field equations for any gauge such as the
models proposed in Ref.~\cite{Zumalacarregui:2013pma,Gleyzes:2014dya,Gleyzes:2014qga, Gao:2014fra, Gao:2014soa}.\\

\section*{Acknowledgments}
The authors are grateful to Ed Copeland for a careful reading of the manuscript and insightful comments. 
The authors acknowledge financial support from  Funda\c{c}\~{a}o para a Ci\^{e}ncia e Tecnologia under the grants EXPL/FIS-AST/1608/2013 and OE/FIS/UI2751/2014. PMM is also supported by the grant PTDC/FIS/111032/2009. FSNL acknowledges financial support by an Investigador FCT Research contract, with reference IF/00859/2012.

\appendix
\section{Dictionary}\label{diccionario}

The Horndeski action can be written as \cite{Horndeski:1974wa} 
\begin{widetext}
\begin{eqnarray}\label{H}
 \mathcal{L}_H&=&\delta^{\alpha\beta\gamma}_{\mu\nu\sigma}\left[\kappa_1\nabla^\mu\nabla_\alpha\phi \,R_{\beta\gamma}{}^{\nu\sigma}
 -\frac{4}{3}\kappa_{1,X}\nabla^\mu\nabla_\alpha\phi\nabla^\nu\nabla_\beta\phi\nabla^\sigma\nabla_\gamma\phi
 +\kappa_3\nabla_\alpha\phi\nabla^\mu\phi\,R_{\beta\gamma}{}^{\nu\sigma}-4\kappa_{3,X}\nabla_\alpha\phi\nabla^\mu\phi\nabla^\nu\nabla_\beta\phi\nabla^\sigma\nabla_\gamma\phi\right]\nonumber\\
 &+&\delta_{\mu\nu}^{\alpha\beta}\left[F\,R_{\alpha\beta}{}^{\mu\nu}-4F_{,X}\nabla^\mu\nabla_\alpha\phi \nabla^\nu\nabla_\beta\phi
 +2\kappa_8\nabla_\alpha\phi\nabla^\mu\phi\nabla^\nu\nabla_\beta\phi\right]
 -3\left[2F_{,\phi}+X\,\kappa_8\right]\nabla_\mu\nabla^\mu\phi+\kappa_9,
\end{eqnarray}
\end{widetext}
where $
 X=\nabla_\mu\phi\nabla^\mu\phi$, $\kappa_i\left(\phi,\,X\right)$ are arbitrary functions, a function $W(\phi)$ which usually appears in (\ref{H}) has been absorbed in $F(\phi,\,X)$ without any lost of generality, 
 and
\begin{equation}\label{condF}
 F_{,X}=\kappa_{1,\phi}-\kappa_3-2X\kappa_{3,X}.
\end{equation}
The functions appearing in the minisuperspace Lagrangian (\ref{Lsimple}) are related with the Horndeski functions by \cite{Charmousis:2011ea})
\begin{eqnarray}
 X_0&=&-\widebar Q_{7,\phi}\dot\phi+\kappa_9,\qquad  X_1=-3\,\widebar Q_7+Q_7\dot\phi,\label{X1}\\
 X_2&=&12\,F_{,X}X-12\,F,\label{X2}\,\,\,\, X_3=8\,\kappa_{1,X}\,\dot\phi^3,\label{X3}\\
Y_0&=&\widebar Q_{1,\phi}\dot\phi+12\,\kappa_3\dot\phi^2-12\,F,\label{Y0}\\
Y_1&=&\widebar Q_1-Q_1\dot\phi\qquad\qquad Y_2=Y_3=0,\label{Y2}
\end{eqnarray}
and we have defined for simplicity (in analogy with Ref.~\cite{Charmousis:2011ea}) the 
functions
\begin{eqnarray}
Q_1&=&\widebar Q_{1,\dot\phi}=-12\,\kappa_1,\label{Q1}\\
Q_7&=&\widebar Q_{7,\dot\phi}=6\,F_{,\phi}-3\,\dot\phi^2\kappa_8.\label{Q7}
\end{eqnarray}
It can be noted that the expression of $X_1(\phi,\,\dot\phi)$ given by Eq.~(\ref{X1}) is not exactly the same as that appearing in \cite{Charmousis:2011ea} because we have simplified it using Eq.~(\ref{Q7}).

As is well known, the Horndeski Lagrangian (\ref{H}) is equivalent to the Lagrangian presented by Deffayet {\it et al.} in Ref.~\cite{Deffayet:2011gz}. This is:
\begin{widetext}
\begin{eqnarray}\label{LDeff}
 \mathcal{L}&=&K(\phi,\,X)-G_3(\phi,\,X)\square\phi+G_4(\phi,\,X)R+G_{4,X}\left[(\square\phi)^2-(\nabla_\mu\nabla_\nu\phi)^2\right]\nonumber\\
 &+&G_5(\phi,\,X)G_{\mu\nu}\nabla^\mu\nabla^\nu\phi-\frac{1}{6}G_{5,X}\left[(\square\phi)^3-3(\square\phi)(\nabla_\mu\nabla_\nu\phi)^2+2(\nabla_\mu\nabla_\nu\phi)^3\right]
\end{eqnarray}
\end{widetext}
The functions appearing in (\ref{H}) and (\ref{LDeff}) are related through the dictionary presented in Refs.~\cite{Deffayet:2011gz} and \cite{Kobayashi:2011nu}. Considering the definitions we are using
this dictionary can be written as
\begin{eqnarray}
 K&=&\kappa_9+X\int^X{\rm d}X'\left(\kappa_{8,\phi}-2\kappa_{3,\phi\phi}\right),\label{K}
\end{eqnarray}
\begin{eqnarray}
 G_3&=&6F_{,\phi}+X\kappa_8+4X\kappa_{3,\phi}\nonumber\\
 &-&\int^X{\rm d}X'\left(\kappa_{8}-2\kappa_{3,\phi}\right),\label{G3}
\end{eqnarray}
\begin{eqnarray}
 G_4&=&2F+2X\kappa_3,\label{G2}
\end{eqnarray}
\begin{eqnarray}
 G_5&=&-4\kappa_1.\label{G1}
\end{eqnarray}
It must be noted that our definition of $X=\nabla_\mu\phi\nabla^\mu\phi$ differs by a factor $1/2$ and a sign to others sometimes used in the literature, what has been taken into account to write
Eqs.~(\ref{K})-(\ref{G1}).\\

\section{Linear models in the notation of Deffayet et al.}\label{funcionesD}
In Sec.~\ref{Requiring} we have presented the Horndeski functions corresponding to the linear models. In this appendix we show the functions that appear in the equivalent Lagrangian (\ref{LDeff})
obtained by Deffayet {\it et al.}~\cite{Deffayet:2011gz}.
Taking into account Eqs.~(\ref{kappa1})-(\ref{Ffinal}) into Eqs.~(\ref{K})-(\ref{G1}), we obtain
\begin{widetext}
\begin{eqnarray}
 K&=& 3\sqrt{\Lambda}U_{0,\phi}+\sqrt{-X}\left(W_{0}-\sqrt{\Lambda}U_{1,\phi}\right)-\frac{X}{6}\left[-3W_{1,\phi}+
  \left(W_{1,\phi}-\sqrt{\Lambda}U_{2,\phi\phi}\right){\rm ln}(-X)\right]\nonumber\\
 &+&(-X)^{3/2}\left[-4U_{4,\phi\phi}+\frac{W_{2,\phi\phi}}{6}\left[-6+{\rm ln}(-X)\right]+\frac{3\sqrt{\Lambda}U_{3,\phi\phi\phi}}{2}
 \left[-2+{\rm ln}(-X)\right]\right]\nonumber\\
 &+&\frac{X^2W_{3,\phi\phi\phi}}{4}\left(-3+{\rm ln}(-X)\right)
\end{eqnarray}
\begin{eqnarray}
 G_3&=&\frac{W_{1,\phi}}{6}\left[-1+{\rm ln}(-X)\right]-\frac{\sqrt{\Lambda}U_{2,\phi}}{2}\left[2+{\rm ln}(-X)\right]
 -8U_{4,\phi}\sqrt{-X}
 +\frac{W_{2,\phi}}{3}\left[-5+{\rm ln}(-X)\right]\sqrt{-X}\nonumber\\
 &+&3\sqrt{\Lambda} U_{3,\phi\phi}\left[1-{\rm ln}(-X)\right]\sqrt{-X}
+\frac{W_{3,\phi\phi}(\phi)X}{4}\left[7-3\,{\rm ln}(-X)\right],
\end{eqnarray}
\begin{eqnarray}
 G_4&=&-\frac{\sqrt{\Lambda}U_2(\phi)}{2}+\sqrt{-X}\left[-2U_4+\frac{W_{2}}{12}\left[-4+{\rm ln}(-X)\right]-
 \frac{3\sqrt{\Lambda}U_{3,\phi}}{4}{\rm ln}(-X)\right]\nonumber\\
 &+&\frac{W_{3,\phi}X}{4}\left[2-{\rm ln}(-X)\right],
\end{eqnarray}

\end{widetext}
\begin{eqnarray} 
 G_5&=&-\frac{3\sqrt{\Lambda} U_3(\phi)}{\sqrt{-X}}+\frac{W_{3}}{2}\,{\rm ln}(-X),
\end{eqnarray}
As it can be noted the form of these functions is not particularly simple, although it can be simpler for particular choices of the functions $U_i(\phi)$ and $W_i(\phi)$.


\begin{thebibliography}{99}



\bibitem{Barcelo:2014qva}
  C.~Barcel\'o, R.~Carballo-Rubio and L.~J.~Garay,
  ``Absence of cosmological constant problem in special relativistic field theory of gravity'',
  arXiv:1406.7713 [gr-qc].

\bibitem{Horndeski:1974wa}
  G.~W.~Horndeski,
  ``Second-order scalar-tensor field equations in a four-dimensional space'',
  Int.\ J.\ Theor.\ Phys.\  {\bf 10} (1974) 363.

\bibitem{Deffayet:2011gz}
  C.~Deffayet, X.~Gao, D.~A.~Steer and G.~Zahariade,
  ``From k-essence to generalised Galileons'',
  Phys.\ Rev.\ D {\bf 84} (2011) 064039
  [arXiv:1103.3260 [hep-th]].

\bibitem{Deffayet:2009wt}
  C.~Deffayet, G.~Esposito-Farese and A.~Vikman,
  ``Covariant Galileon'',
  Phys.\ Rev.\ D {\bf 79} (2009) 084003
  [arXiv:0901.1314 [hep-th]].

\bibitem{Nicolis:2008in}
  A.~Nicolis, R.~Rattazzi and E.~Trincherini,
  ``The Galileon as a local modification of gravity'',
  Phys.\ Rev.\ D {\bf 79} (2009) 064036
  [arXiv:0811.2197 [hep-th]].

\bibitem{deRham:2014zqa} 
  C.~de Rham,
  ``Massive Gravity'',
  Living Rev.\ Rel.\  {\bf 17}, 7 (2014)
  [arXiv:1401.4173 [hep-th]].

\bibitem{Charmousis:2011bf}
  C.~Charmousis, E.~J.~Copeland, A.~Padilla and P.~M.~Saffin,
  ``General second order scalar-tensor theory, self tuning, and the Fab Four'',
  Phys.\ Rev.\ Lett.\  {\bf 108} (2012) 051101
  [arXiv:1106.2000 [hep-th]].
  
\bibitem{Charmousis:2011ea}
  C.~Charmousis, E.~J.~Copeland, A.~Padilla and P.~M.~Saffin,
  ``Self-tuning and the derivation of a class of scalar-tensor theories'',
  Phys.\ Rev.\ D {\bf 85} (2012) 104040
  [arXiv:1112.4866 [hep-th]].

\bibitem{Copeland:2012qf}
  E.~J.~Copeland, A.~Padilla and P.~M.~Saffin,
  ``The cosmology of the Fab-Four'',
  JCAP {\bf 1212} (2012) 026
  [arXiv:1208.3373 [hep-th]].

\bibitem{Appleby:2012rx}
  S.~A.~Appleby, A.~De Felice and E.~V.~Linder,
  ``Fab 5: Noncanonical Kinetic Gravity, Self Tuning, and Cosmic Acceleration'',
  JCAP {\bf 1210} (2012) 060
  [arXiv:1208.4163 [astro-ph.CO]].
  
\bibitem{Linder:2013zoa}
  E.~V.~Linder,
  ``How Fabulous Is Fab 5 Cosmology?'',
  JCAP {\bf 1312} (2013) 032
  [arXiv:1310.7597 [astro-ph.CO]].

\bibitem{Gubitosi:2011sg}
  G.~Gubitosi and E.~V.~Linder,
  ``Purely Kinetic Coupled Gravity'',
  Phys.\ Lett.\ B {\bf 703} (2011) 113
  [arXiv:1106.2815 [astro-ph.CO]].

\bibitem{Martin-Moruno:2015lha}
  P.~Mart\'{\i}n-Moruno, N.~J.~Nunes and F.~S.~N.~Lobo,
  ``Attracted to de Sitter: cosmology of the linear Horndeski models'',
  arXiv:1502.05878 [gr-qc].

\bibitem{Heisenberg:2014kea}
  L.~Heisenberg, R.~Kimura and K.~Yamamoto,
  ``Cosmology of the proxy theory to massive gravity'',
  Phys.\ Rev.\ D {\bf 89} (2014) 103008
  [arXiv:1403.2049 [hep-th]].

\bibitem{DeFelice:2010pv}
  A.~De Felice and S.~Tsujikawa,
  ``Cosmology of a covariant Galileon field'',
  Phys.\ Rev.\ Lett.\  {\bf 105} (2010) 111301
  [arXiv:1007.2700 [astro-ph.CO]].

\bibitem{Zumalacarregui:2013pma}
  M.~Zumalac\'arregui and J.~Garc\'{\i}a-Bellido,
  ``Transforming gravity: from derivative couplings to matter to second-order scalar-tensor theories beyond the Horndeski Lagrangian'',
  Phys.\ Rev.\ D {\bf 89} (2014) 6,  064046
  [arXiv:1308.4685 [gr-qc]].

\bibitem{Gleyzes:2014dya}
  J.~Gleyzes, D.~Langlois, F.~Piazza and F.~Vernizzi,
  ``Healthy theories beyond Horndeski'',
  arXiv:1404.6495 [hep-th].

\bibitem{Gleyzes:2014qga}
  J.~Gleyzes, D.~Langlois, F.~Piazza and F.~Vernizzi,
  ``Exploring gravitational theories beyond Horndeski'',
  arXiv:1408.1952 [astro-ph.CO].

\bibitem{Gao:2014soa}
   X.~Gao,
   ``Unifying framework for scalar-tensor theories of gravity'',
   Phys.\ Rev.\ D {\bf 90} (2014) 8, 081501
   [arXiv:1406.0822 [gr-qc]].
   
\bibitem{Gao:2014fra}
   X.~Gao,
   ``Hamiltonian analysis of spatially covariant gravity'',
   Phys.\ Rev.\ D {\bf 90} (2014) 10, 104033
   [arXiv:1409.6708 [gr-qc]].

\bibitem{Kobayashi:2011nu}
  T.~Kobayashi, M.~Yamaguchi and J.~'i.~Yokoyama,
  ``Generalized G-inflation: Inflation with the most general second-order field equations'',
  Prog.\ Theor.\ Phys.\  {\bf 126} (2011) 511
  [arXiv:1105.5723 [hep-th]].

\end{thebibliography}
\end{document}